\documentclass{llncs}
\usepackage{grffile}
\usepackage{graphicx}
\usepackage{comment}
\usepackage{amsmath,amssymb} 
\usepackage{color}
\usepackage{makecell}

\usepackage[width=122mm,left=12mm,paperwidth=146mm,height=193mm,top=12mm,paperheight=217mm]{geometry}

\begin{document}
\pagestyle{headings}
\mainmatter
\def\ECCVSubNumber{5479}  

\title{Efficient Unpaired Image Dehazing with Cyclic Perceptual-Depth Supervision} 

\titlerunning{Unpaired Image Dehzing with Depth Supervision}
\author{Chen Liu\inst{\star}\and
Jiaqi Fan\inst{\star} \and
Guosheng Yin}

%
%
\institute{The University of Hong Kong, Pokfulam, Hong Kong \email{\{liuchen,garyfan,gyin\}@hku.hk}}
\maketitle
\footnotetext[1]{The two authors contributed  equally to this paper.}
\begin{abstract}
Image dehazing without paired haze-free images is of immense importance, as acquiring paired images often entails significant cost. However, we observe that previous unpaired image dehazing approaches tend to suffer from performance degradation near depth borders, where depth tends to vary abruptly. 
Hence, we propose to anneal the depth border degradation in unpaired image dehazing with cyclic perceptual-depth supervision. Coupled with the dual-path feature re-using backbones of the generators and discriminators, our model achieves $\mathbf{20.36}$ Peak Signal-to-Noise Ratio (PSNR) on NYU Depth V2 dataset, significantly outperforming its predecessors with reduced Floating Point Operations (FLOPs).
\end{abstract}

\section{Introduction}

Image dehazing has wide practical implications in poor weather condition autonomous driving, visibility estimation, and photography. Various image dehazing approaches, majority of which requiring paired images, have been developed in the literature, from utilizing image statistics~\cite{he2010single,Berman_2016_CVPR} to modeling haze with convolutional neural networks~\cite{morales2019feature,zhang2018densely}.

Given the difficulty in collecting haze-free counterparts to hazy images, there have been growing interests in unpaired image dehazing in recent years. Unpaired image deahazing negates the necessity of having a matching haze-free image, thereby significantly reducing the cost of data acquisition. Engin et al.~\cite{engin2018cycle} first adopted CycleGAN~\cite{CycleGAN2017} to generate haze-free images from cycle-consistency constraints, followed with extensions by Zhao et al.~\cite{zhao2019dd} and Dudhane et al.~\cite{dudhane2019cdnet}. However, we identify two areas in the previous work~\cite{engin2018cycle} that calls for improvements: first, it tends to have trouble removing haze at regions near depth borders; second, it lacks mechanisms to model the relationship between haze and scene depth.  

To resolve the two issues as above, we propose to use a dual-path feature re-using backbone inspired by Octave Convolutions~\cite{Chen_2019_ICCV} for the generators and discriminators of CycleGAN, to model the low complexity scenes with lower resolution paths and exploit details with higher resolution paths.

We summarize our major contributions as follows:
\begin{itemize}
    \item Empirically, we demonstrate that perceptual-depth supervision helps the model to learn a more realistic mapping.
    \item By disentangling feature maps of different spatial frequencies, we strengthen the representation power and reduce the computational cost, resulting in an efficient and high performance unpaired image dehazing network.  
\end{itemize}

\begin{figure}[!]
\centering
\includegraphics[width=0.8\linewidth]{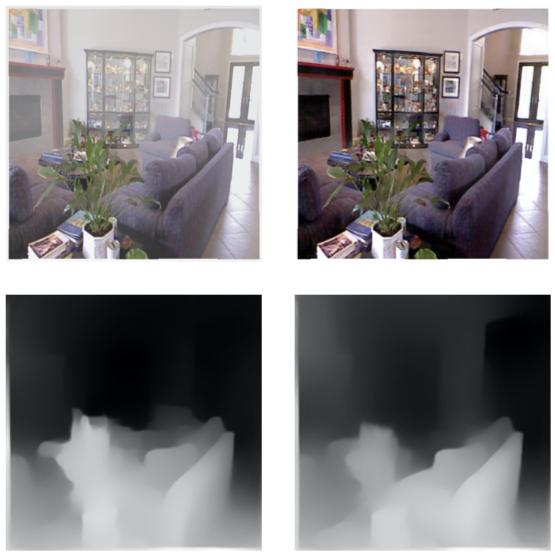}
\caption{The first row: a sample hazy image and the dehazing results on the NYU Depth V2 training set. The model is equipped with Octave Convolution backbone and cyclic perceptual depth loss, trained to the 125th epoch. The second row: depth maps obtained from a pre-trained monocular depth estimation network~\cite{lasinger2019towards}. } 
\label{fig:monodepth}
\end{figure}

\section{Related Work}

\subsection{Atmospheric Scattering Model}

The presence of haze in images is often formulated by the atmospheric scattering model~\cite{zhang2018densely,narasimhan2000chromatic} as 

\begin{equation}\label{eq:scatter}
    \mathcal{I}\left(x\right) = \mathcal{J}\left(x\right) t\left(x\right)  + A(1 - t\left(x\right) ) ,
\end{equation} 

where $  \mathcal{I}\left(x\right)$ is the hazy image, the input, $\mathcal{J}\left(x\right)$ is the clear image, the output, $t\left(x\right)$ is the transmission map, and $A$ is the global atmospheric light. The transmission map is related to the scene depth $d$ in the form of 
\begin{equation}\label{eq:depth}
    t\left(x\right) = \exp\left(-d\beta\left(\lambda\right)\right) ,
    \end{equation} 

where $d$ is the distance between the camera sensor and the target scene. $ \beta(\lambda)$ represents the total scattering coefficient, referring to the ability of a volume to scatter flux of a given wavelength $\lambda$ in every direction.

\subsection{Unpaired Image-to-Image Translation}
Zhu et al.~\cite{CycleGAN2017} proposed to map the distribution of $X$ to that of $Y$ with a generator $G_A: X \rightarrow Y$. To learn such a mapping without paired images, they introduced a cycle-consistency loss, which imposes that given another generator $G_B: Y \rightarrow X$ that aims to learn an inverse mapping, $\hat X = G_B(G_A(X)) \approx X$. Formally, the loss can be written as 
\begin{equation}\label{eq:cycle_consistency_loss}
\begin{split}
        \mathcal{L}_{cyc}\left(G_A, G_B \right) &=  \mathop{\mathbb{E}}_{x \sim P_{data}(x)} \left\| G_B\left(G_A\left(x \right) \right) -  x  \right\|_1 \\
    &+ \mathop{\mathbb{E}}_{\hat{x} \sim P_{data}(\hat x)} \left\| G_A\left(G_B\left(\hat{x}\right)\right) -  \hat{x}  \right\|_1 .
\end{split}
\end{equation}

This approach has achieved great success in style transfer and domain adaptation.


\subsection{Octave Convolution}

\begin{figure}[h]
\centering
\includegraphics[width=0.8\linewidth]{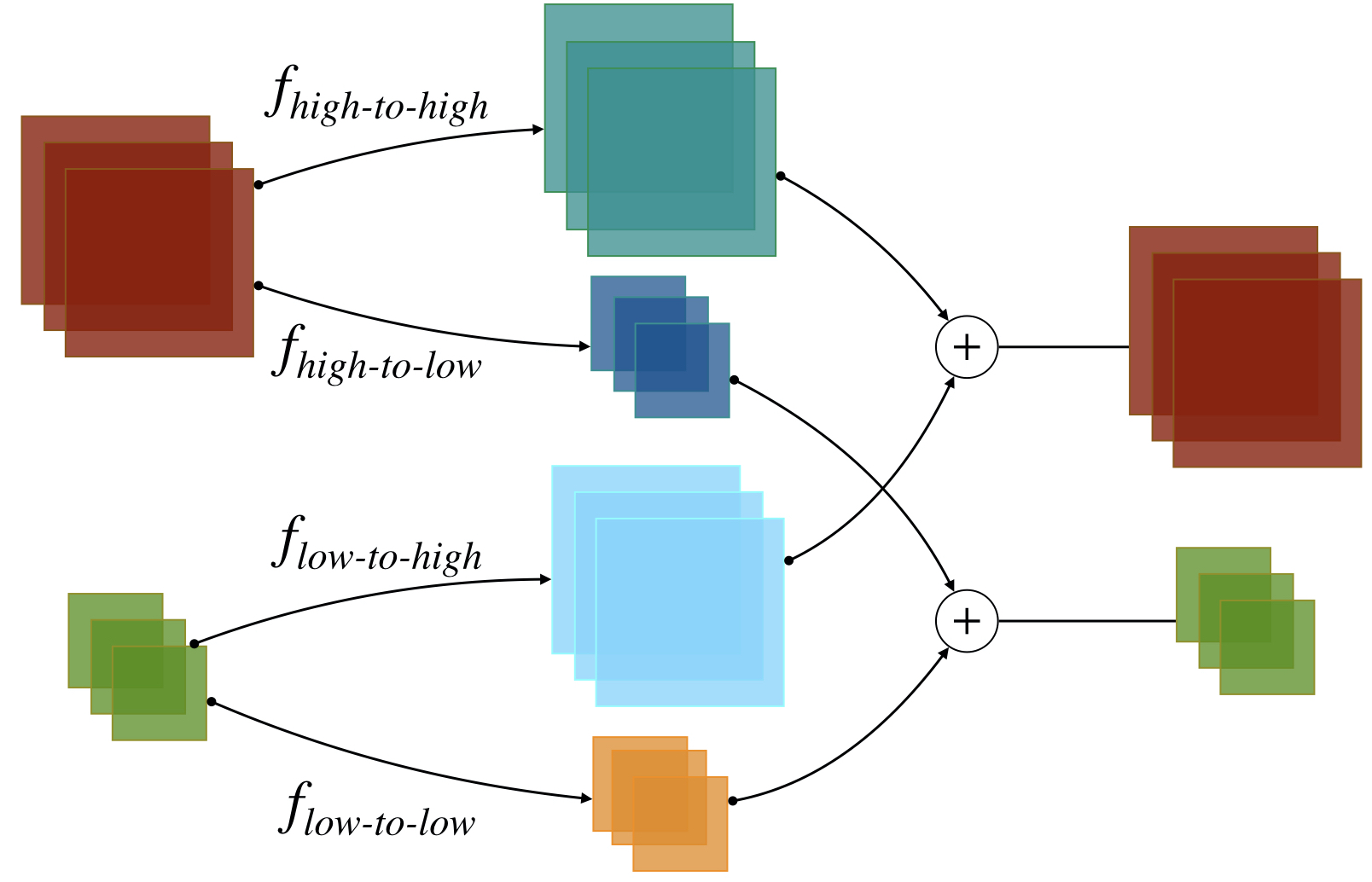}
\caption{The illustration of Octave Convolutions, where $\oplus$ denotes the element-wise sum. Note that in practice, the high resolution feature maps and low resolution feature maps may have different channel dimensions. }
\label{fig:octave}
\end{figure}

Octave Convolutions~\cite{Chen_2019_ICCV} separately model the high and the low spatial frequency image signals by two sets of convolution operations, as illustrated in Fig.~\ref{fig:octave}.

Each set of convolution operations consists of two convolution filters without sharing parameters. The first set of the convolution operations takes the higher resolution input feature map and generates two outputs: one retaining the spatial resolution while the other halving the spatial resolution. The second set of the convolution filters takes the lower resolution input feature maps and also generates two outputs: one having the spatial resolution unchanged and the other having its spatial resolution up-scaled by 2 times. In effect, the Octave Convolution layer maps two input feature maps at different spatial dimensions into four feature maps, two at a higher spatial resolution and another two at a lower spatial resolution. Taking element-wise sums of those feature maps at the same-dimension gives the final outputs. As the summands are from the high spatial resolution input and the low resolution input, lateral inter-spatial resolution communication is achieved.  

\begin{figure}[h]
\begin{center}
\includegraphics[width=0.8\linewidth]{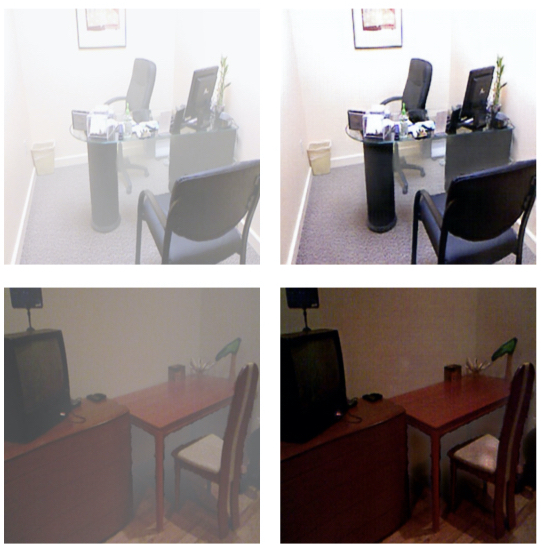}
\end{center}
\caption{Example of the baseline unpaired image dehazing results. The left column consists of hazy training images, and the right column corresponds to the dehazed outputs from a Cycle-dehaze~\cite{engin2018cycle} model trained on NYU Depth V2 to 223 epochs (upper right) and 220 (lower right) epochs. Note that despite the prolonged training schedule, the model still struggles at the region where depth varies dramatically, e.g. residual haze leftover underneath the desk on the lower right image.}
\label{fig:exampledepthborder}
\end{figure}

\section{Method}

For paired images dehazing, a recent trend is to adopt multi-scale learning~\cite{Liu_2019_ICCV} and iterative processing~\cite{liu2019learning}, both of which, however, can be computationally intensive or memory costly.  

Furthermore, they do not take the depth into account. Fig.~\ref{fig:exampledepthborder} highlights the importance of the depth awareness in the dehazing model, as it shows a depth-unaware dehazing model struggling to deal with hazy images with tricky depth variation, resulting in images that are often haze-free in areas with smooth depth variations but hazy in areas where the depth varies abruptly. 

To introduce depth awareness in to our model, we leveraged the observation that the effectiveness of the depth estimation algorithms can be adversely affected by the haze present in images. Fig.~\ref{fig:monodepth} shows an example where the depth estimation algorithm failed to capture the hallway in the background to the right of the image due to the hazy environment. As such, the predictions from depth estimation algorithms, sensitive to depth variations, can be utilized as supervision signals. Incorporated into the dehazing model, it serves to introduce the depth awareness. 
\begin{figure}[h]
\centering
\includegraphics[width=1.0\linewidth]{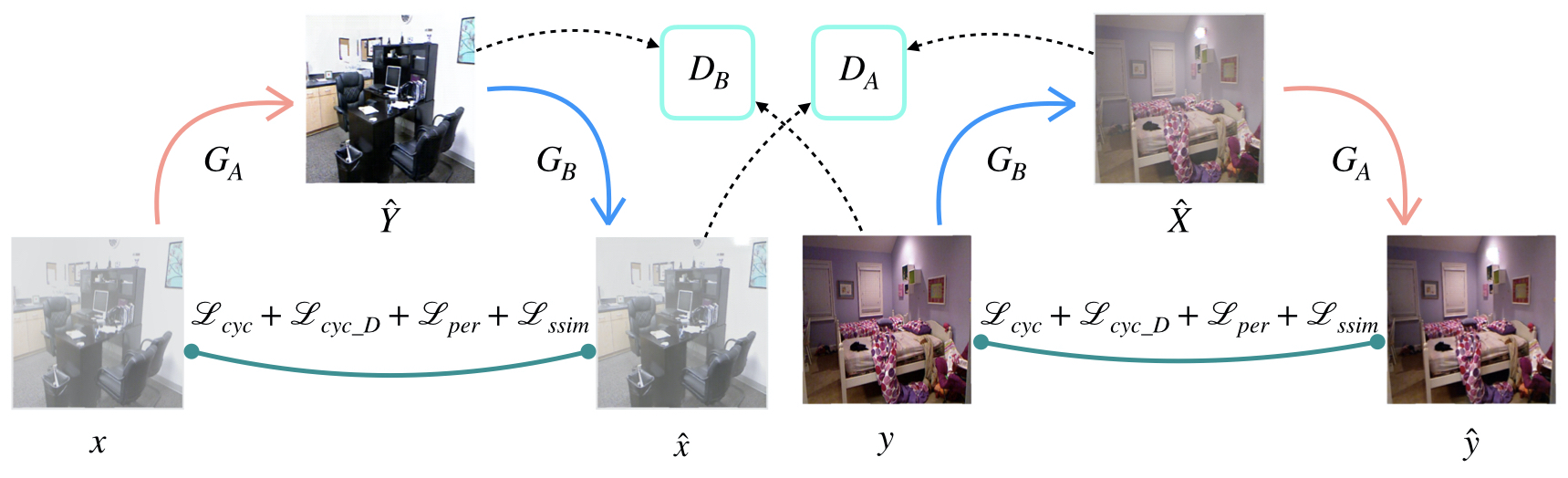}
\caption{The overall structure of the proposed dehazing model. $D_B$ aims to distinguish between $\hat Y$ and $y$, corresponding to the generated dehazed images and real world haze-free images. Note that the coefficients of different losses are omitted for illustration purposes. The identity mapping losses are also omitted since they are present in all experimental configurations.}
\label{fig:overall_arch}
\end{figure}

We built our approach on top of CycleGAN and incorporated a pre-trained monocular depth estimation network by Xian et al.~\cite{xian2018monocular} as a Cyclic perptual loss component. Adding to the depth-aware model is Octave Convolutions for enhanced feature representation learning. The overall architecture of the proposed model is illustrated in Fig.~\ref{fig:overall_arch}.

\subsection{Enhanced Feature Representation in GAN} 

We adopt Octave Convolutions~\cite{Chen_2019_ICCV} in both the generators and discriminators to simultaneously model the fine details around depth borders and the coarse features in regions with smooth depths of low variations. Self-attention modules  \cite{vaswani2017attention} are also tested for modeling inter-pixel dependencies.

\paragraph{\textbf{Octave Convolution in generators.}} The generators follow the implementations in Johnson et al.~\cite{Johnson_2016} and Zhu et al.~\cite{CycleGAN2017}, with the backbones redesigned with Octave Convolution blocks between the downsampling and upsampling layers. The output feature maps of the downsampling layers, also the input to subsequent Octave Convolution layers, are split in the channel dimension into two, with high resolution and low resolution, respectively. This also reduces the computational complexity. 

\paragraph{\textbf{Octave Convolution in discriminators.}} Regular 3-layer networks with Octave Convolutions form the discriminators. As half of the channels in the Octave Convolutions are equipped with enlarged receptive field, the discriminators achieve better global context awareness with fewer parameters. 
In the ablation studies, we show empirically that this design choice leads to significant performance boost.

\paragraph{\textbf{Self-attention modules.}} Following the configurations in Zhang et al.~\cite{zhang2018selfattention}'s work, we incorporate the self-attention into the backbones of the generators and discriminators. Table~\ref{table:ablation} shows that it brings sizeable performance benefits, albeit unlike the octave convolution, it comes at the cost of added parameters.

\subsection{Cyclic Perceptual Depth-consistency Loss}
Based on (\ref{eq:scatter}) and (\ref{eq:depth}), the transmission map for a given hazy image is a function of the distance between the camera sensor and location. However, as only one hazy RGB image is given as input for the dehazing task, the scene depth is under-constrained through the transformation. Apart from the cycle consistent loss, and cyclic perceptual and cyclic perceptual-consistency loss, we argue that the learned map functions should also preserve depth information: For an image $x$, from the hazy image domain $X$, the two mapping functions, $G$ and $F$, should be able to restore the scene depth feature of $x$. Likewise, after cyclic transformation, the output image with an image $y$ from the clear image domain as an input image should share similar scene depth features.

We define the perceptual depth-consistency loss that measures depth differences between the original and reconstructed images, which utilizes a scene depth loss network $\phi$. In all our experiments, $\phi$ is the  ResNet-based multi-scale architecture proposed by Xian et al.~\cite{xian2018monocular} for monocular depth estimation, which is pretrained on the 3D Movies dataset by Lasinger et al.~\cite{lasinger2019towards}. 

The perceptual depth-consistency loss is represented as the squared L2 norm between two scene depth maps, 
\begin{equation}\label{eq:depth_loss}
    \mathcal{L}_{D}\left(x, \hat{x} \right) =  \left\| \phi\left(x\right) -  \phi\left(\hat{x}\right)  \right\|_2^2 . 
\end{equation}

As two transformation directions should be given equal consideration, the cyclic perceptual depth-consistency loss can be formulated as
\begin{equation}\label{eq:cyclic_depth_loss}
    \mathcal{L}_{Cyc\_D}=  \mathcal{L}_{D}\left(x, F\left(G\left(x\right)\right)\right) +  \mathcal{L}_{D}\left(y,G\left(F\left(x\right)\right) \right) . 
\end{equation}


\subsection{Implementations}
This section introduces how different modules are integrated into the network as well as training details.

\paragraph{\textbf{Integration of enhanced feature representation modules.}}
To strike a balance between model complexity and performance, all Octave Convolution layers in the experiments have equal number of channels in their lower spatial frequency branch and higher spatial frequency branch. 
Where to add the feature enhancement modules is a non-trivial question. Following \cite{Chen_2019_ICCV}, in the Octave Convolution experiments, the initial downsampling layers of generators and discriminators and the upsampling layers in the generators are left unchanged for stable training. As there is no relative position embedding in the self-attention, we found completely replacing residual blocks with self-attention modules degrades the performance. Instead, self-attention modules are added to the last 2 of the residual blocks in the generators and all but the first convolution layers in the discriminators. 

\paragraph{\textbf{Training details.}}
The baseline are performed on $512\times512$ images resized to $256\times256$ with two images per batch. Adam optimizer with initial learning rate $0.0002$ and linear decay schedule was used throughout the experiments. No data augmentation is performed apart from horizontal flipping. No random crop is performed as it easily loses global context and is found to jeopardize the performance. 


\section{Experiments}
This section first introduces the datasets. Next, we report thorough ablation studies on various components of our model. We then compare the performance of the proposed method with state-of-the-art unpaired single image dehazing quantitatively and qualitatively. 

\subsection{Datasets} 
\paragraph{NYU Depth V2 dataset.} NYU Depth V2 consists of a wide range of complex images in the RGBD format, collected from various indoor locations in the US cities~\cite{nyudepthsilberman2012indoor}. We utilized the datasets prepared by Zhang and Patel on the NYU Depth V2 dataset for training and validation~\cite{zhang2018densely}. Specifically, 4000 training samples are synthesized from 1000 unique images according to (\ref{eq:scatter}). The total scattering coefficient $\beta$ and global atmospheric light $A$ of the training set are randomly and uniformly selected. The sampling process ensures that the hazy level of the datasets follows a continuous uniform distribution with parameters $A$ and $\beta$. Similarly, a testing dataset consisted of 400 samples is synthesized from another 100 images in NYU Depth V2. 
\begin{table}[t]
\begin{center}
\begin{tabular}{cc|c|cc|c}
\Xhline{1.0pt}
\multicolumn{6}{c}{(a) \textbf{Backbone architecture}}                              \\
G & D & Loss &PSNR & SSIM &  \#param \\
\hline  
9B & 3L & base &   16.01 & 0.68  &  28.29M \\ 
6B-SA & 3L-SA & base & 17.63 & 0.73 & 22.35M  \\ 
6B-Oct & 3L & base & 17.62 & 0.73 & 21.47M  \\ 
\Xhline{1.0pt}
\multicolumn{6}{c}{(b) \textbf{Convolution block design}}  \\
G & D & Loss & PSNR & SSIM &  \#param \\
\hline
6B-Oct &  3L  & CPD  & 18.08 & 0.75 & 21.47M \\ 
6B-Oct &  3L-Oct &  CPD  &  20.25 & 0.79 & 19.06M  \\  
6B-Oct &  3L-OctN &  CPD & \textbf{20.36} & \textbf{0.80} &20.14M\\ 
\Xhline{1.0pt}
\multicolumn{6}{c}{(c) \textbf{Additional loss components}}\\
G & D & Loss &PSNR & SSIM &  \#param \\
\hline
6B-Oct &  3L  & base & 17.62 & 0.73 & 21.47M  \\ 
6B-Oct &  3L  &  CPD  & 18.08 & 0.75 & 21.47M \\   
6B-Oct &  3L  &  SSIM  & 17.45 & 0.73 & 21.47M \\ 
6B-Oct &  3L-Oct  &  CPD  &  20.25 & 0.79 &19.06M  \\ 
6B-Oct &  3L-Oct  & SSIM  & 18.99 &  0.75 & 19.06M  \\ 
6B-Oct &  3L-OctN & CPD & \textbf{20.36} & \textbf{0.80} & 20.14M\\ 
6B-Oct &  3L-OctN & SSIM & 20.01 & 0.79 & 20.14M  \\ 
\Xhline{1.0pt}
\end{tabular}
\end{center}
\caption{\textbf{Ablation study} based on CycleGAN with 9 residual blocks in the generator and 3 layer patch GAN discriminator. ``SA" refers to self-attention; Column ``G" and ``D" refer to the architecture of generators and discriminators, respectively. ``6B" and ``9B" in Column `` G" refers to six and nine residual blocks in generator, respectively. ``Oct" in Column ``G" and ``D" refers to OctConv. ``3L" means 3 layer. ``N" in Column `` D" refers to the adoption of SpectralNorm in discriminator. Column ``Loss" indicated the additional loss components apart from basic loss functions employed by almost all experimental configurations, namely cycle-consistency loss, identity mapping loss, and perceptual ResNet loss. The ``base" in Column ``Loss" refers to basic loss functions mentioned above. ``CDP" refers to \textbf{C}yclic \textbf{P}erceptual \textbf{D}epth-\textbf{C}onsistency Loss.  }
\label{table:ablation}
\end{table}

\subsection{Ablation study on NYU Depth V2 dataset}
We explored three components that affect the performance of unsupervised single image dehazing: 
\begin{itemize}
    \item overall backbone structure, 
    \item convolution block design, 
    \item the additional loss components, including the structural similarity (SSIM) index, and cyclic perceptual depth-consistency and  total variation loss.
\end{itemize}
The ablation study based on CycleGAN is done on the NYU Depth V2 testing dataset. PSNR and SSIM are reported to evaluate model performance. Additionally, the number of parameters is included in Table~\ref{table:ablation} to evaluate the computational cost.

\paragraph{\textbf{Overall backbone structure.}}
Following~\cite{engin2018cycle}, we begin with using the CycleGAN with 9 residual blocks in the generator and a 3-layer patch GAN discriminator as the baseline. Then, we replace the 9 residual blocks between the downsampling and upsampling operations with 6 residual blocks and one self-attention layer between the last two blocks. Meanwhile, a 3-layer discriminator is used, including two self-attention layers among the convolution filters. Next, We use 6 Octave Convolution blocks to substitute the residual blocks in the generator of the baseline model but remain the regular 3-layer patch GAN discriminator. All three backbones are trained for 200 epochs with cyclic consistency loss and ResNet perceptual loss. As shown in Table~\ref{table:ablation}(a), by fewer parameters and less computation, the backbone with Octave Convolution achieves performance comparable to that incorporated with self attention.  
\paragraph{\textbf{Convolution block design.}}
To find the optimal integration scheme of Octave Convolution and the backbone of CycleGAN, we investigate the effectiveness of OctConv in the discriminator and the generator. In Table~\ref{table:ablation}(b), we compare the performance of two models with the same octave generator but with three different discriminators: (1) regular 3-layer, (2)  the 3-OctConv-layer, and (3) the 3-OctConv-layer with Spectral Normalization~\cite{miyato2018spectral}. All three models are trained with cyclic consistency and depth-consistency loss. We observe a significant improvement as convolution filters in the discriminator are substituted by Octave Convolution(12.00\% increase on PSNR, 5.33 \% increase on SSIM). By further adding SpectralNorm to the discriminator yields an increase.  Hence, we adopt the Octave Convolution generator and Octave SpectralNorm discriminator. Instead of VGG16 perceptual loss~\cite{Johnson_2016}, we calculated the perceptual loss a concatenation of all 5 layer outputs from a ImageNet~\cite{krizhevsky2012imagenet} pretrained ResNet50~\cite{he2016deep}.     

\paragraph{\textbf{Additional loss components.}} Apart from the cyclic consistency, identity mapping, and perceptual ResNet loss, we thoroughly tested cyclic depth-consistency loss, Cyclic SSIM loss on different combinations of backbones and modules. In this paper, SSIM loss is defined as 
\begin{equation}\label{eq:ssim_loss}
    \mathcal{L}_{SSIM}\left(x, \hat{x} \right) = 1 - \mathbf{SSIM}\left(x, \hat{x} \right) ,
\end{equation}

where $\mathbf{SSIM}$ follows the definition in~\cite{ssimwang2004image}. 
Thus, the cyclic SSIM loss is formulated as 
\begin{equation}\label{eq:cyclic_ssim_loss}
    \mathcal{L}_{Cyc\_SSIM}= \mathcal{L}_{SSIM}\left(x, F\left(G\left(x\right)\right)\right)  + \mathcal{L}_{SSIM}\left(y,G\left(F\left(x\right)\right) \right) .
\end{equation}

As illustrated in Table~\ref{table:ablation}(c), adding cyclic depth-consistency loss yields higher performance (6.64 \% increase on PSNR, 8.22 \% increase on SSIM) with a slight increase of FLOPs on the backbone with the Octave Convolution generator and discriminator, while adding SSIM loss caused a minor drop on PSNR. Furthermore, on the other two backbones, adding cyclic depth-consistency loss always results in improvement compared with adding SSIM loss. 
The consistent performance gain on a variety of backbone confirms that cyclic depth-consistency loss is a functional loss component for dehazing problems.

\subsection{Comparing with unpaired dehazing SOTAs}
This section first lays out the difference in our re-implementation of Cycle-dehaze to the original and then compares our proposed model to the Cycle-dehaze on NYU Depth V2.
We re-implemented the original Cycle-dehaze in Pytorch with identical 9 residual blocks in the generators and VGG16~\cite{simonyan2014very} preceptual loss. The original Cycle-dehaze trained on NYU Depth V2 for around 40 epochs~\cite{engin2018cycle}, while both our model and our re-implementation is trained for 200 epochs. Our model, reported in Table~\ref{table:sota}, adds the cyclic perceptual-depth loss with a pretrained monocular depth estimation network from~\cite{lasinger2019towards}, adopts Octave Convolution\cite{Chen_2019_ICCV} in the generators and the discriminators, and uses Spectral Normalization~\cite{miyato2018spectral} solely in the discriminators. One critical difference is that our model reported is equipped with only 6 Octave Convolution blocks, instead of 9 in the Cycle-dehaze model (both the original and re-implementation), while still achieves significantly better result. 
\begin{table}
\begin{center}
\begin{tabular}{|l|c c|}
\hline
Method & PSNR & SSIM\\
\hline\hline
Cycle-dehaze Re-implement & 16.15 & 0.657  \\
Cycle-dehaze Claimed & 15.41  & 0.66 \\
Octave \& Depth & 20.36 & 0.80 \\
\hline
\end{tabular}
\end{center}
\caption{Results on NYU Depth V2, note that there are notable differences in our re-implementation of Cycle-dehaze, elaborated below. }
\label{table:sota}
\end{table}

\subsection{Qualitative Results}
\begin{figure}[!b]
\begin{center}
\includegraphics[width=0.95\linewidth]{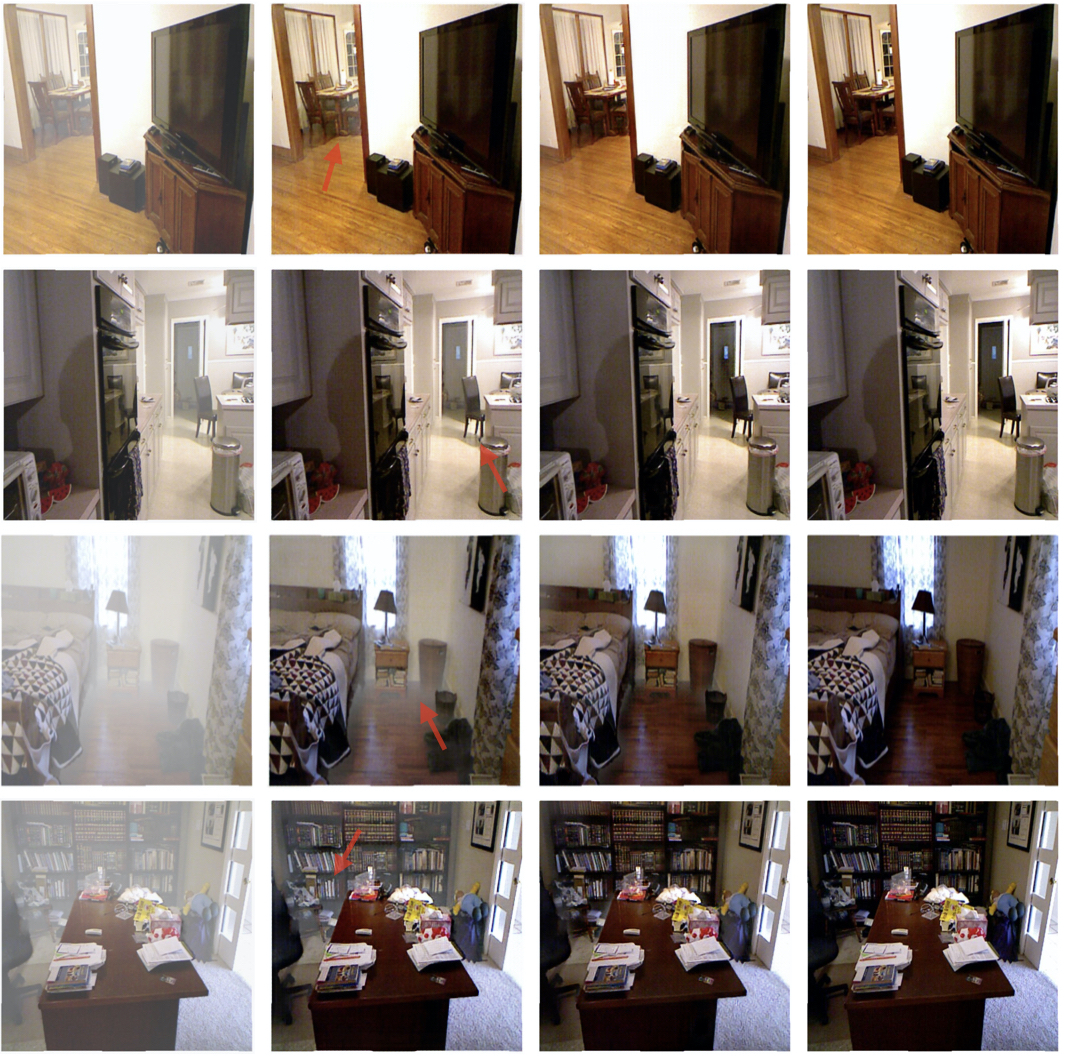}
\end{center}
\caption{Sample comparison between CycleGAN with ResNet50 perceptual loss and our model with cyclic perceptual depth-consistency loss and octave convolution. Column-wise, From left to right are synthesized hazy images, dehazed outputs from CycleGAN, dehazed outputs from our model, and real haze-free images. Pictures taken from the test set of NYU Depth V2.}
\label{fig:quali}
\end{figure}

In this section, we qualitatively compare the performance of our proposed method with CycleGAN. The CycleGAN model is trained with cycle-consistency loss, adversarial training loss, identity mapping loss functions, and ResNet50 perceptual loss. Our efficient unpaired image dehazing model is trained with cyclic perceptual-depth supervision and Octave Convolution. Both models are trained and validated on NYU Depth V2 dataset for 200 epochs. Results on NYU Depth V2 testing set are presented in Fig.~\ref{fig:quali}. Note that our model handles the challenging depth border regions better.

\section{Conclusions}

This paper draws attention to the issues present in existing unpaired image dehazing approaches and proposed a light-weight and efficient unpaired image dehazing network through cycle-consistency adversarial training and cyclic perceptual-depth supervision. Our proposed cyclic perceptual-depth consistency loss improves the consistency in generated images with negligible computational overhead and does not slow down the inference. We also adopt Octave Convolution to formulate a disentangled two-path feature representation backbone, which, added to the discriminators, yields significant improvement over the traditional designs.  



\clearpage
%
%
\bibliographystyle{unsrt}

\end{document}